\documentclass{nature}

\usepackage{graphicx}
\usepackage{amssymb}
\usepackage{hyperref}
\usepackage{xcolor}
\usepackage{amsmath}
\usepackage{textcomp}

\newcommand{\NewEdits}[1]{{{\color{black}#1}}}

\author{ P.\ Shah$^{1,\dagger}$, S.\ Arora$^{1,\dagger}$, M.M.\ Driscoll $^{1,*}$}

\title{Coexistence of solid and liquid phases \NewEdits{in shear jammed} colloidal drops}

\begin{document}

\maketitle

\begin{affiliations}
 \item Department of Physics and Astronomy, Northwestern University
 \item[$\dagger$] These authors contributed equally to this work.
 \item[*] corresponding author
\end{affiliations}

\section*{Abstract}

Complex fluids exhibit a variety of exotic flow behaviours under high stresses, such as shear thickening and shear jamming. Rheology is a powerful tool to characterise these flow behaviours over the bulk of the fluid. However, this technique is limited in its ability to probe fluid behaviour in a spatially resolved way. Here, we utilise high-speed imaging and the free-surface geometry in drop impact to study the flow of colloidal suspensions. We report, for the first time, observations of coexisting solid and liquid phases \NewEdits{due to shear jamming caused by impact}. In addition to observing Newtonian-like spreading and bulk shear jamming, we observe the transition between these regimes in the form of localised patches of jammed suspension in the spreading drop. We capture shear jamming \textit{as it occurs} via a solidification front travelling from the impact point, and show that the speed of this front is set by how far the impact conditions are beyond the shear thickening transition. 

 \section{Introduction}
Complex fluids, such as particulate suspensions\cite{Wagner2012colloidal,StickelPowellRheo} and polymer solutions\cite{osswald2015polymer}, exhibit a variety of exotic flow behaviours, for instance shear thickening and solidification via jamming. These behaviours are particularly relevant to development of smart materials, such as body armours\cite{DavidArmor} and soft robots\cite{rus2015design}. Rheometry is traditionally used to characterise complex fluids. However, this technique typically provides measurements averaged over the bulk of the fluid and obscures the information on local variations in flow. The free-surface geometry in drop impact systems offers a unique lens to probe these flow properties, as it provides data on manifestations of non-Newtonian flow \NewEdits{with high spatial and temporal resolution.} Here, we use high-speed imaging to study the drop impact of colloidal suspensions over a large range of volume fractions and impact velocities, thus sampling impact behaviour from liquid-like spreading to solid-like jamming. Combined with input from rheological data, our measurements offer a more holistic picture of complex fluid flow\NewEdits{, especially under dynamic conditions}. 
 
An extensive understanding has been developed for the dynamics of a Newtonian fluid drop impacting a dry solid substrate\cite{josserand2016drop,philippi_2016_newtonian,laan2014maximum, lee_laan_bonn_2016, gordillo_force_2018,ChengStressReview}. However, the vastly different flow properties of complex fluids substantially modify impact dynamics. Past studies have explored the spreading and splashing of a variety of \NewEdits{polymeric fluids\cite{Bergeron2000, blackwell2015sticking} and particulate suspensions\cite{peters2013splashing,thoraval2021nanoscopic,boyer2016drop, jorgensen2020deformation,bertola2015impact,kim2019impact}}; however each has largely focused on a relatively narrow slice of the vast parameter space. \NewEdits{The role of particulate additives in controlling the splashing transition has been explored\cite{peters2013splashing, thoraval2021nanoscopic}, as well as the spreading and jamming of dense suspension drops\cite{boyer2016drop, jorgensen2020deformation, bertola2015impact, kim2019impact}. In particular, experiments on impacting shear thickening fluids have reported solid-like states after impact\cite{boyer2016drop,jorgensen2020deformation, bertola2015impact}}.  Colloidal suspensions offer a convenient control parameter --- volume fraction --- to scan suspension behaviours ranging from Newtonian-like to shear thickening. Here, we report a systematic study of colloidal suspension impact spanning a range of volume fractions ($0.09 \le \phi \le 0.50$) and impact velocities ($0.7$ $ m/s \le u_0 \le 4.0$ $m/s$). \NewEdits{Our exploration of this wide parameter space allows us to capture not only spreading and bulk solidification, but also the transition between these drastically different flow regimes.}

\NewEdits{Shear thickening --- an increase in viscosity with increasing shear --- is one of the most counterintuitive phenomena exhibited by dense suspensions\cite{Wagner2012colloidal, StickelPowellRheo}. The Peclet number, $Pe=\frac{\text{shear rate}}{\text{rate of diffusion}}$, is a dimensionless number used to quantify high shear rates, and we expect the onset of shear thickening at $Pe \gg 1$. The transition to shear thickening occurs when a suspension with relatively high $\phi$ is subjected to a shear higher than a critical value\cite{Wagner2012colloidal, BrownJaeger2014shear, DennMortonSoftMatter, MorrisThickening2020}. We note that for our experimental parameters, $Pe>10^2$, and thus we expect shear thickening after impact in the high-$\phi$ limit.   Many rheological studies have focused on elucidating the mechanism of shear thickening, and both lubrication hydrodynamics and particle interactions have been shown to play a role. Shear thickening has been proposed as precursor to shear jamming, and the nature of this transition is an active field of study.  For a more detailed discussion, we refer the reader to the following reviews\cite{StickelPowellRheo, BrownJaeger2014shear,DennMortonSoftMatter, MorrisThickening2020}.}
 
For our experiments, we synthesize  charge-stabilized silica spheres (diameter 830 $\pm$ 20 nm, Fig.\ \ref{rheology}a)  using the St\"ober process\cite{stober1968,zhang2009} and suspend them in water. Spherical drops of diameter $d_0= 3.0 \pm 0.1$ mm are formed by drawing a known volume of fluid ($15$ \textmu L) into a micropipette. We set the impact velocity by changing the height from which the drops are released, and record the drops impacting on a dry, hydrophilic glass substrate using a high-speed camera. To minimize the effects of particle sedimentation, all samples are re-suspended immediately before the experiments using a vortex mixer. All experiments are performed in a humidity chamber, which additionally mitigates air currents (see Methods for details).




To connect impact behaviours with rheological properties, a mapping between impact velocity and rheological parameters such as shear rate or shear stress is necessary. Precisely quantifying shear rates in drop impact systems is challenging due to the nonuniformity of shear in both space and time. However, a simple dimensional argument can be used to estimate the shear rate at impact. At the instant of impact, the bottom point of the drop comes to rest, while the apex continues to fall at the impact velocity $u_0$, as the \NewEdits{shear caused by} impact has not had time to propagate \NewEdits{across the drop}. Dividing this difference in speeds, $u_0$, by the drop size $
d_0$ thus provides an estimate of the maximum shear rate at the moment of impact: $\dot{\gamma}_{impact}=u_0/d_0$. With the drop size of $3$ mm, we could access shear rates in the range $233$ $s^{-1}\le \dot \gamma_{impact} \le 1333$ $s^{-1}$. Thus, we are able to span a large spectrum of flow behaviours in these suspensions, and observe how non-Newtonian flow \NewEdits{gives} rise to a rich variety of impact phenomena. The results we present here take us closer to an understanding of the shear jamming transition and the properties of shear jammed solids.

 \section{Results}

\begin{figure}
\begin{centering}
\includegraphics[width=\textwidth]{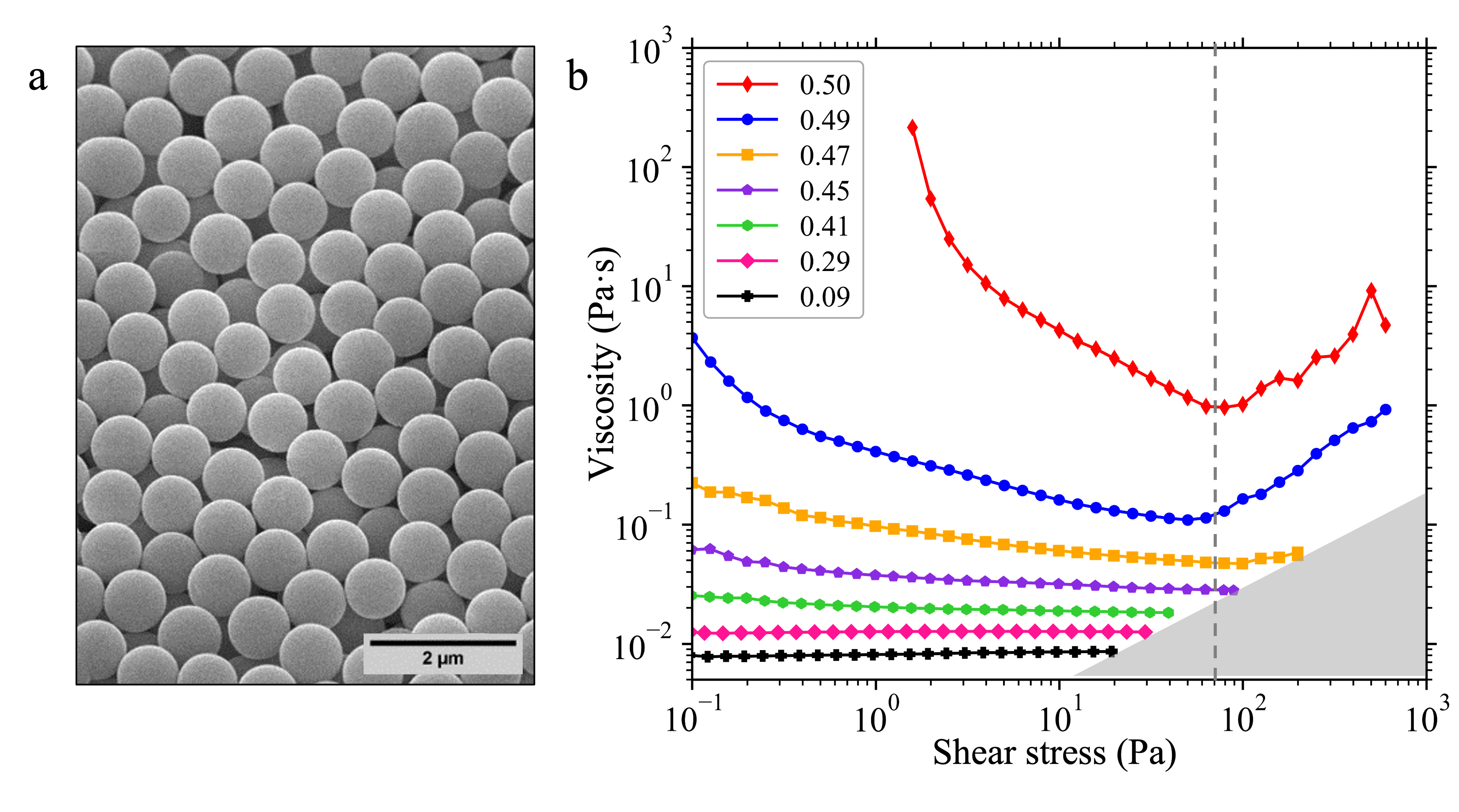} \caption{\textbf{Rheology of the colloidal suspensions.} 
\textbf{a} SEM image of the colloidal silica spheres used in our drop impact experiments; the sphere diameter is 830 $\pm$ 20 nm. \textbf{b} Bulk rheological flow curves: the colloidal suspension exhibits viscous flow, shear thinning, and shear thickening as $\phi$ is increased. The grey triangle in the bottom right indicates the rate limit of the rheometer.}
\label{rheology}
\end{centering}
\end{figure}

\begin{figure}
\begin{centering}
\includegraphics[width=0.95\textwidth]{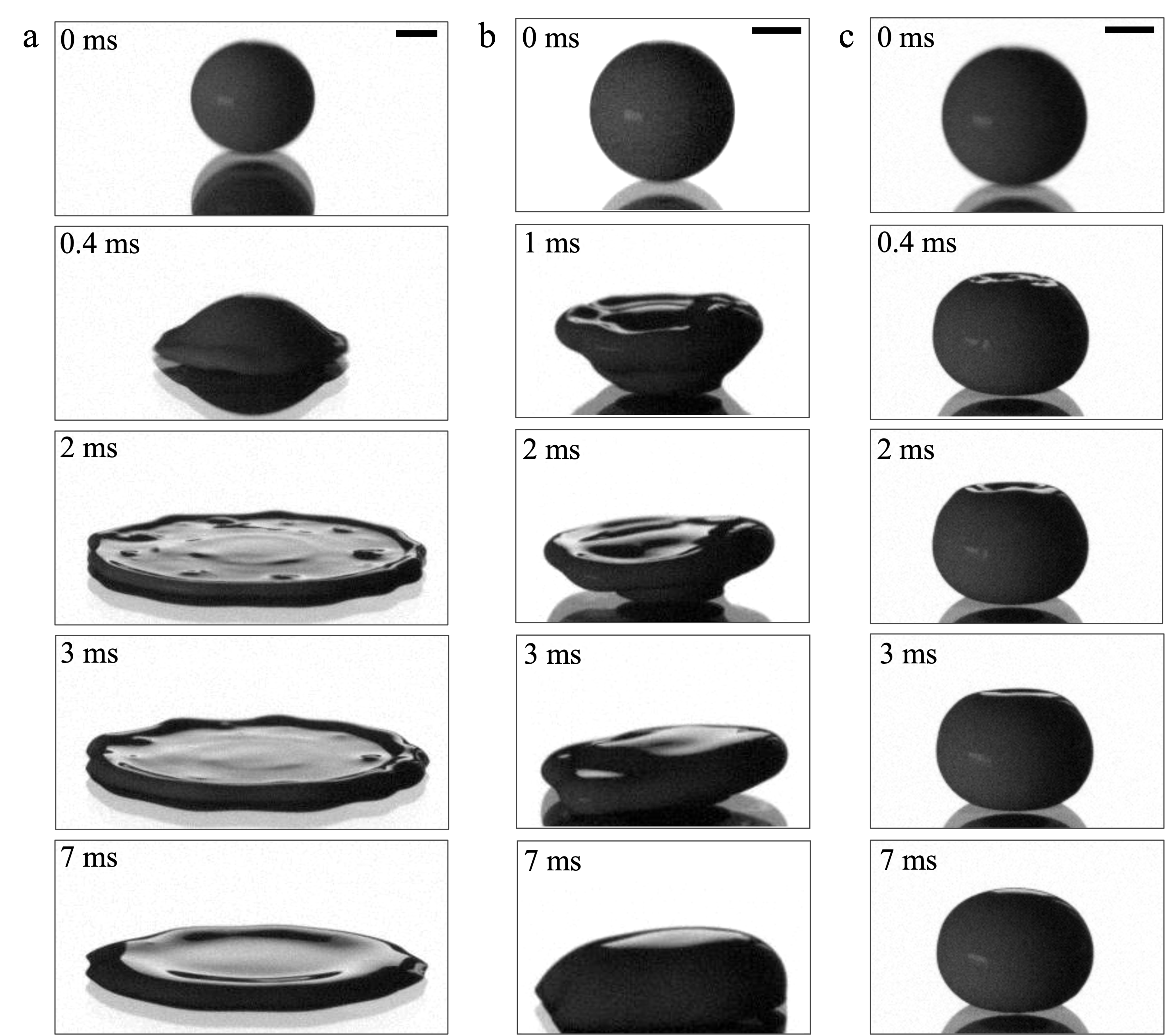} \caption{\textbf{Exotic impact behaviours of colloidal suspension drops.} 
\textbf{a} Timeseries of a $\phi=0.47$ colloidal drop expanding after impacting at $u_0=$ 3.0 m/s [SI video 1 \cite{SI}]. The spreading drop shows transient pockets of localized solidification, indicating the onset of shear thickening.  \textbf{b} Timeseries of a  $\phi=0.49$ colloidal drop impacting at $u_0=$ 2.0 m/s [see also SI video 2]. The bottom half of the drop solidifies, while the still-fluid top portion flows over it. \textbf{c} Timeseries of a  $\phi=0.49$ drop impacting at $u_0=$ 3.0 m/s [SI video 3 \cite{SI}]. While most of the drop is solidified, the top portion of the drop is in the liquid phase. All scale bars are 1 mm.}

\label{pictures}
\end{centering}
\end{figure}

Bulk rheometry measurements [Fig.\ \ref{rheology}b] demonstrate the variety of flow behaviours exhibited by our suspensions.  At low $\phi$ (black and pink lines), the fluid viscosity is constant, akin to a Newtonian fluid. Shear thinning (decreasing viscosity) becomes pronounced as $\phi$ is increased (green and purple curves), and shear thickening (indicated by increasing viscosity) appears for $\phi\ge$ 0.47 at high shear stresses (\NewEdits{shear stress above 100 Pa}, orange, blue, and red curves). We observe fascinating consequences of this non-Newtonian rheology in our impact experiments. At $\phi=0.47$, where weak shear thickening appears at high stresses in bulk rheology, we observe patches of localised solidification during spreading --- panel 3 of Fig. \ref{pictures}a shows small solid-like bumps that protrude from the spreading drop, but vanish in panels 4 and 5 [SI video 1\cite{SI}]. At higher $\phi$, we observe partial solidification of the drop ---   Panel 2 in Fig.\ \ref{pictures}b shows that the bottom part of the drop acts as a solid, while the top part remains fluid and flows over the solidified region  throughout panels 3-5 [SI video 2\cite{SI}]. Finally, at $\phi=0.49$ and high impact velocities, most of drop solidifies as shown in Fig.\ \ref{pictures}c [SI video 3\cite{SI}]. Here, we show that this variety of solidification behaviours is a direct consequence of shear jamming\cite{bi2011jamming}, evidenced by their occurrence much below the static jamming threshold\cite{james2019controlling}.

\begin{figure}
\begin{centering}
\includegraphics[width=\textwidth]{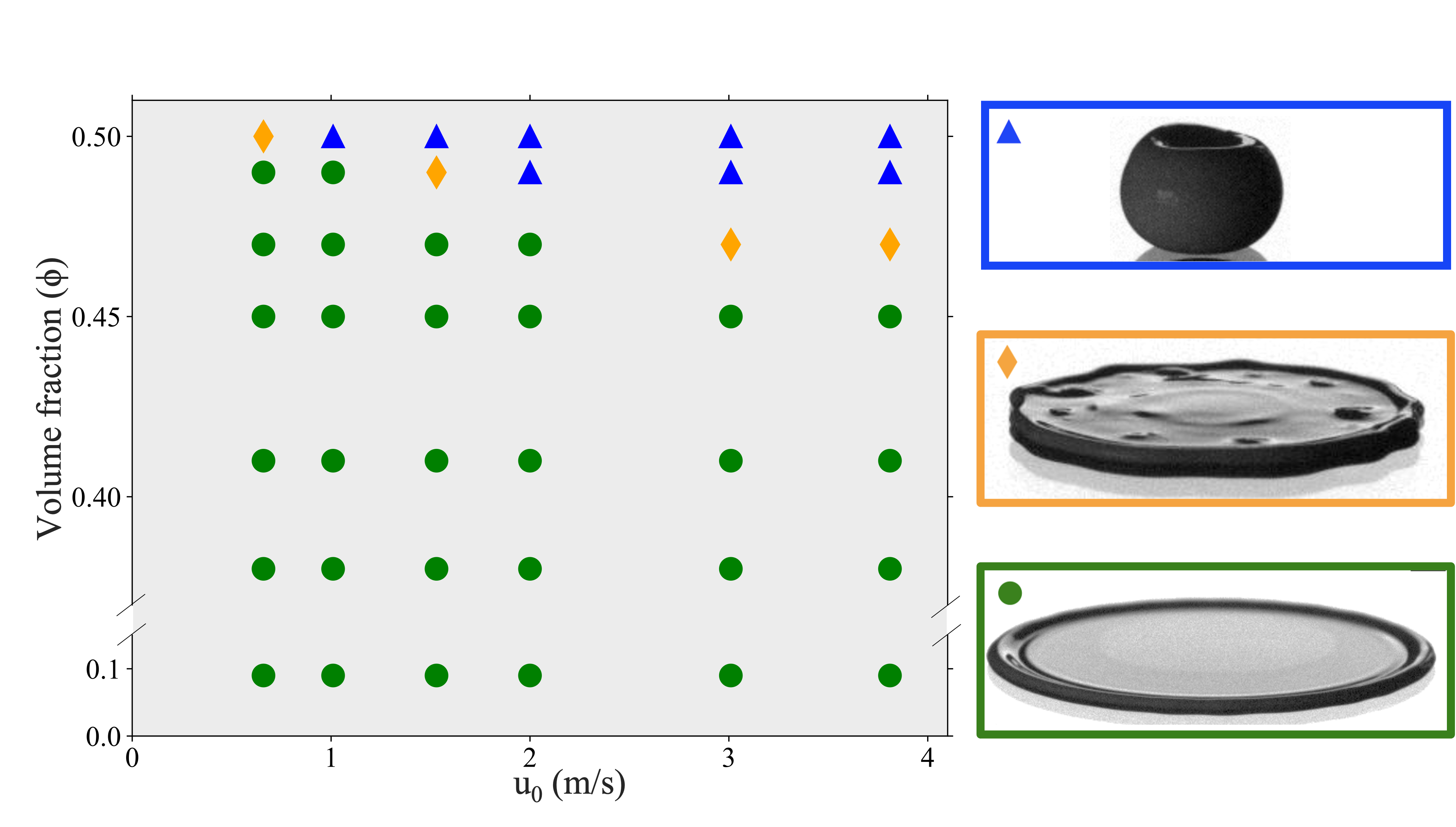} \caption{\textbf{State diagram of colloidal drop impact.} $\phi-u_0$ state diagram summarizing impact regimes; representative snapshots corresponding to these regimes are shown on the right. Green circles denote simple spreading behaviour, which dominates the low $\phi$, low $u_0$ region. Orange diamonds indicate that transient pockets of localised solidification were observed during spreading. Blue triangles correspond to the partial/full solidification regime, where the bottom portion of the drop jams after impact, but a shrinking region at the top remains fluid.
}
\label{statediag}
\end{centering}
\end{figure}

We encapsulate this broad range of impact outcomes in a $\phi-u_0$ state diagram [Fig.\ \ref{statediag}]. Green circles, indicating simple spreading [SI video 4\cite{SI}], dominate the low $\phi$ and low $u_0$ region. With increasing $\phi$ or $u_0$, the localised solidification regime appears (orange diamonds), followed by the bulk solidification regime (blue triangles), where a larger and larger portion of the drop solidifies upon impact. The transition between these regimes is a function of both $\phi$ and $u_0$, as all regimes can be accessed by varying either of the parameters while keeping the other constant. Additionally, we find that the drop behaviour is very sensitive to small changes in $\phi$, consistent with the transition to shear thickening in rheological measurements [Fig.\ \ref{rheology}b].

\begin{figure}
\begin{centering}
\includegraphics[width=\textwidth]{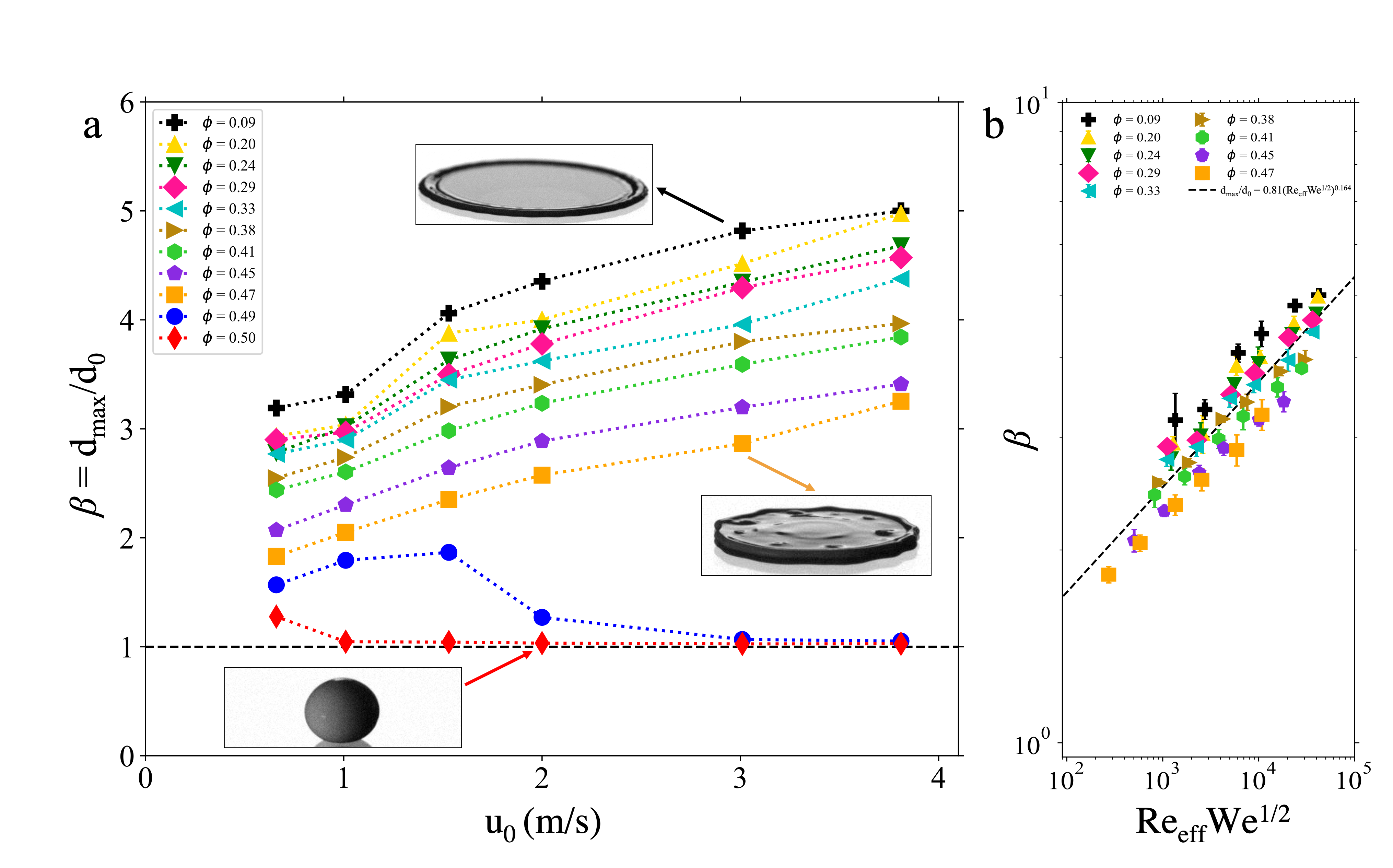} \caption{\textbf{Quantifying \NewEdits{maximum drop spreading.}} ~\textbf{a} Normalised maximum diameter, $\beta=d_{max}/d_{0}$, as a function of $u_0$ for various volume fractions $\phi$.  For $\phi \ge 0.49$ and high impact velocities,  $\beta$ drops to 1, indicating the drop does not spread. Insets show representative snapshots of simple spreading (upper), localised solidification (middle), and bulk solidification (lower). Dotted lines are guides to the eye, and the dashed black line indicates $\beta=$ 1.  \textbf{b} Normalised maximum diameter, $\beta$, for $\phi\le$ 0.47 plotted against the dimensionless parameter $Re_\textit{\scriptsize{\mbox{eff}}}We^{1/2}$. The exponent of the power law fit (dashed line, $\beta=0.81(ReWe^{1/2})^{0.164}$) is in agreement with the scaling reported for Newtonian fluids\cite{Scheller1995}.}
\label{beta}
\end{centering}
\end{figure}

To quantify this range of impact outcomes, we compute the normalized maximum diameter of the impacted drops, $\beta=d_{max}/d_0$, and plot this metric against $u_0$ [Fig.\ \ref{beta}a]. For $\phi\le$ 0.47, $\beta$ increases with increasing impact velocity. However, $\beta$ drops to 1 at $\phi \ge $  0.49 and high impact velocities. This is because the drop no longer spreads after impact (lower inset). This result is consistent with recent studies that observed similar solidification in suspension impact at high $\phi$\cite{boyer2016drop,bertola2015impact}. Our drops remain solid for a few milliseconds; however, they spread like a liquid over the timescale of a second [SI video 5\cite{SI}]. Thus, the solid-like state we observe is transient in nature, further evidence that this solidification is a direct result of shear jamming. A recent result suggests that the substrate wettability affects this timescale of unjamming\cite{bertola2015impact}, but this problem remains largely unexplored.    

At $\phi\le$ 0.47, the drops spread in a manner qualitatively  similar to Newtonian fluids [SI video 4\cite{SI}]. Previous experiments\cite{Scheller1995} with Newtonian fluids have shown that $\beta$ scales as the dimensionless parameter $ReWe^{1/2}$,  where $We$ is the impact Weber number, $\rho u_0^2d_0/\sigma$, and $Re$ is the Reynolds number, $Re=\rho u_0d_0/\eta$. \NewEdits{Here, $\sigma$ is the surface tension of the suspending fluid (for this case water, $\sigma=72$ mN/m), $\rho$ is the fluid density, calculated as:}
\NewEdits{
\begin{equation}
\rho=\rho_{silica}\phi+\rho_{water}(1-\phi)
\end{equation}
}
\NewEdits{ with $\rho_{silica}=2000$ kg/m$^3$ and $\rho_{water}= 1000$ kg/m$^3$, and $\eta$ is the suspension viscosity.} For impacting colloidal drops, the calculation for the Weber number remains identical to Newtonian fluids. Estimating the Reynolds number, however, is less straightforward due to the non-constant fluid viscosity of complex fluids. Immediately after impact, the shear rate experienced by the drop is at its maximum ($\dot\gamma_{impact}=u_0/d_0$). As the fluid spreads and slows down, the shear rate continuously drops to zero. We use the average of these two extremes, $\dot\gamma_{avg}=u_0/2d_0$, as the most straightforward estimate of shear rate throughout the spreading process. We then use the viscosity value corresponding to  $\dot\gamma_{avg}$ from our bulk rheological data to calculate the effective Reynolds number, $Re_\textit{\scriptsize{\mbox{eff}}}$ at each $\phi$ and $u_0$. \NewEdits{For our experimental conditions, the range of dimensionless numbers was $20<We<1000$ and $50<Re_\textit{\scriptsize{\mbox{eff}}}<1600$.} In Fig.~\ref{beta}b, we plot $\beta$ in the spreading regime against $Re_\textit{\scriptsize{\mbox{eff}}}We^{1/2}$; the dashed black line indicates the power-law fit:

\begin{equation}
\beta=\NewEdits{(0.81\pm0.02)} (Re_\textit{\scriptsize{\mbox{eff}}}We^{1/2})^{\NewEdits{0.164\pm0.002}}.
\end{equation}
The exponent of this fit agrees with that reported by Scheller et al.\cite{Scheller1995} for viscous Newtonian fluids, $\beta=0.61(ReWe^{1/2})^{0.166}$. Thus, our simple estimate of shear rate and in turn effective viscosity, provides a useful framework to quantify the maximum spread of impacting colloidal drops. \NewEdits{Given that the fluids considered in this study are highly shear thinning at higher $\phi$, the agreement we report here with Newtonian models is surprising. We note that more recent works\cite{laan2014maximum,lee_laan_bonn_2016} have modified the scaling by Scheller et al.\cite{Scheller1995} to  additionally account for surface wettability. However, incorporating these effects for impacting colloidal fluids is highly non-trivial, and would likely modify only the low-$We$ spreading behaviour. }

In the localised solidification regime [orange diamonds in Fig.\ \ref{statediag}, SI video 1\cite{SI}] the bulk of the drop still spreads like a Newtonian fluid [Fig.\ \ref{beta}b], but shear thickening is apparent via solidified patches that appear and then disappear. \NewEdits{These patches appear during the spreading phase, around 1 millisecond after impact. However, indicators of jamming are present earlier, in the form of nonuniformity in the spreading rim of the drop [see for example, panel 2 in Fig.\ \ref{pictures}a]. In most cases, these patches outlive the spreading phase and disappear during the receding phase, over tens of milliseconds.} Our observation of this regime coincides with the onset of weak shear thickening in the bulk rheology data [orange curve in Fig.\ \ref{rheology}b]. Moreover, the transient nature of these patches is strong evidence that regions of high viscosity are embedded in a lower-viscosity fluid phase. \NewEdits{We note that we can only observe these patches on the drop surface in high-speed imaging data, and there is a large variance in the spatial and temporal distribution of these patches. This limits our ability to extract quantitative information about localised jamming.}  For higher $\phi$, where shear thickening is pronounced, the drop \NewEdits{exhibits drastically different behaviour, and does not spread at all.}

\begin{figure}
\begin{centering}
\includegraphics[width=\textwidth]{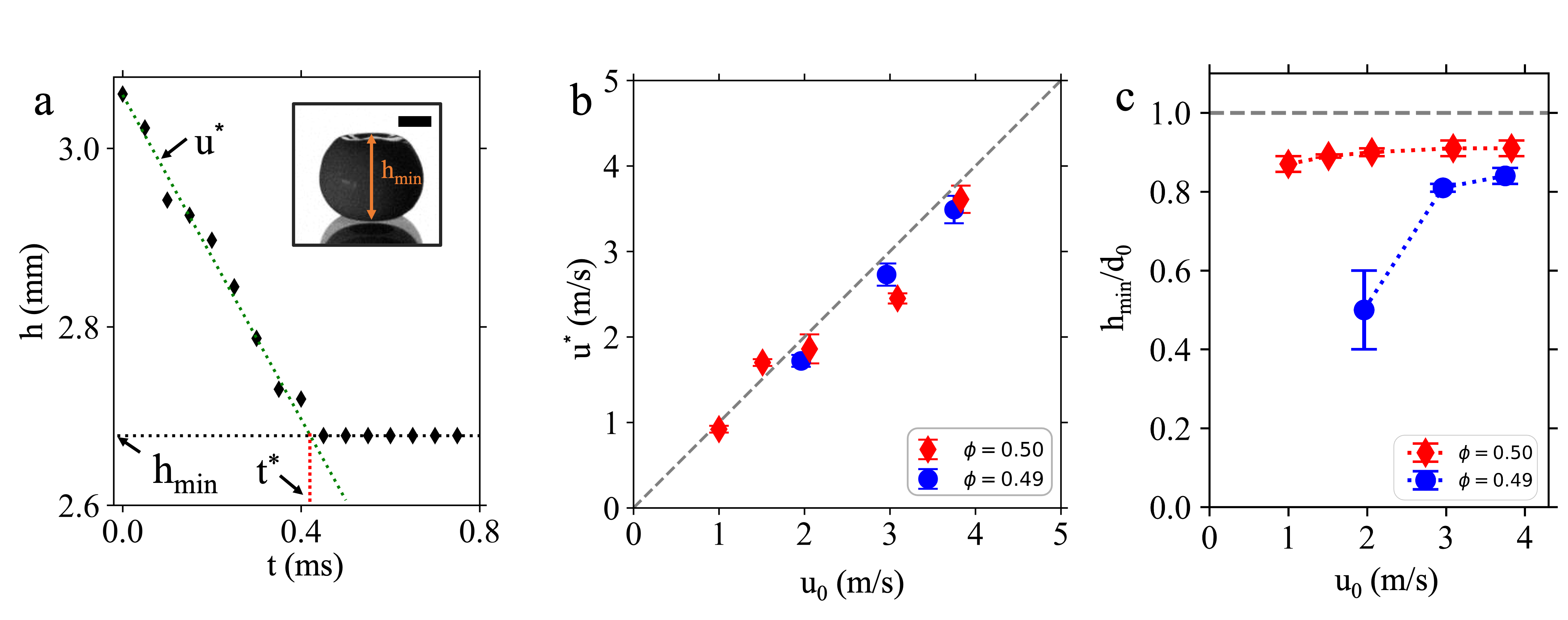} \caption{\textbf{Characterisation of the partial solidification regime.} ~\textbf{a} Height of the \NewEdits{drop apex} from the impact substrate, plotted against time, for $\phi=0.50$, $u_0=3$ $m/s$. $h$ decreases at the speed $u^\ast$ until time $t^\ast$, then plateaus at the value $h_{min}$. Inset: post-impact snapshot of a drop at the minimum height $h_{min}$. \textbf{b} $h$ decreases at a speed identical to the impact velocity, indicating that over the timescale $t^\ast$, the top portion of the drop is unaffected by the impact event. Dashed line corresponds to $u^\ast=u_0$. \textbf{c} $h_{min}/d_0$ vs.\ impact velocity $u_0$. $h_{min}/d_0$ increases with increasing impact velocity, and then plateaus at a value less than 1, indicating finite compression of the drop along impact axis. Dashed line indicates $h_{min}/d_0=1$.}
\label{solid}
\end{centering}
\end{figure}

For $\phi\ge$ 0.49, a large fraction of the drop solidifies upon impact. To quantify the dynamics of this partially solidified state, we measure the height of the \NewEdits{drop apex} as a function of time [Fig.\ \ref{solid}a]. Consistent with another study of impacting shear-thickening drops\cite{boyer2016drop}, we observe two regimes in the $h$ vs.\ $t$ curve \NewEdits{--- a free-fall regime and a plateau regime}. Immediately after impact, $h$ decreases at a rate identical to the impact velocity \NewEdits{(free-fall regime) [Fig.\ \ref{solid}b]}, and then plateaus at a constant value, $h_{min}$ \NewEdits{(plateau regime)}. This is strong evidence that any shear from the impact event has not yet propagated to the top portion of the drop, and hence the top portion must still remain a liquid. \NewEdits{Studies of impacting Newtonian drops have also observed a similar `free-fall' regime where the drop apex moves at the impact velocity\cite{lagubeau_Newtonian_2012,gordillo_force_2018,mitchell_Newtonian_2019}. It is worth noting that in contrast to Newtonian fluids, where a broad transition regime was observed between the free-fall and the plateau regimes, we observe a sudden transition from the free-fall to plateau regime [Fig.\ \ref{solid}a], a direct indication of a shear jammed drop. }

We quantify the spatial extent of solidification by plotting the normalized minimum height, $h_{min}/d_0$ against $u_0$ [Fig.\ \ref{solid}c]. The increase in $h_{min}/d_0$ with $u_0$ indicates that a larger and larger volume of the drop is solidified  as the impact velocity is increased. Interestingly, at high impact velocities, $h_{min}/d_0$ plateaus to a value smaller than $1$, indicating that the solidified drop also undergoes \NewEdits{deformation} along the impact direction, \NewEdits{along with slight bulging in the plane transverse to impact [Fig.\ \ref{pictures}c]}. Furthermore, the high temporal resolution (100,000 fps) of our imaging enables us to capture the details of this solidification as it occurs.

Immediately after impact, we observe a disturbance travelling upward along the drop surface over hundreds of microseconds [orange and green arrows in Fig.\ \ref{front}a]. 
To better visualise this front, we subtract successive frames of the image sequence, so that only the parts that change between frames are highlighted [right panel of Fig.\ \ref{front}a, SI video 6\cite{SI}]. The location of the front is given by the lower end of the bright edge [Fig.\ \ref{front}b]. As this front travels upward, the portion of the drop above the front still maintains its pre-impact curvature [red circles in Fig.\ \ref{front}a], indicating that it is unaffected by the impact event until the front reaches it (consistent with $u^\ast=u_0$, Fig.\ \ref{solid}b). The angular location of this front plotted against time reveals that the front travels at a constant speed, $u_\textit{\scriptsize{\mbox{front}}}$ [slope of the line in Fig.\ \ref{front}c]. $u_\textit{\scriptsize{\mbox{front}}}$ increases with increasing $u_0$, and its value is several times larger than $u_0$ [Fig.\ \ref{front}d]. As evident from the rheology, the suspension thickens when the applied shear surpasses a critical value. Indicated by the dotted lines in Fig.\ \ref{front}e, the critical shear rate where thickening is observed, $\dot\gamma_c$, is much lower for $\phi$ = 0.50 than for $\phi$ = 0.49. We plot $u_\textit{\scriptsize{\mbox{front}}}$ against the excess shear rate over this critical value, $\dot\gamma_{impact}-\dot\gamma_c$, and the data indeed collapses on a single curve for both $\phi$ [Fig.\ \ref{front}f] . \NewEdits{This suggests that} the speed of this disturbance is set by this excess shear rate. 



As the impact velocity is increased, a larger and larger volume of the drop solidifies upon impact. At $\phi=0.50$ and $u_0=4$ $m/s$, we observe that the drop \NewEdits{bounces} off the substrate, with the coefficient of restitution $\epsilon=0.1$ [SI video 7\cite{SI}]. \NewEdits{This rebound behaviour is especially striking given the hydrophilic nature of the substrate.}  By coupling this coefficient of restitution with the drop's \NewEdits{deformation} along the impact axis, we can semi-empirically estimate the elastic modulus of the solidified drop. The drop impacts the substrate with an initial velocity $u_0$, remains in contact with the substrate for time $\Delta t=200$ \textmu s , and then rebounds with the final velocity $\epsilon u_0$. While in contact with the substrate, we measure that the drop is \NewEdits{deformed} \NewEdits{in the direction of impact} by the amount $\Delta x=0.24$ mm. 
We calculate the force experienced by the drop upon impact using momentum conservation:
\begin{equation}
F=\frac{m\Delta u}{\Delta t}=\frac{m(1+\epsilon)u_0}{\Delta t},
\end{equation}
To convert the force to a stress, we divide by the contact area for a Hertzian contact\cite{landau1986theory}, $\pi a^2=\pi d_0\Delta x/2$:
\begin{equation}
\sigma=\frac{F}{\pi d_0\Delta x/2}=\frac{2m(1+\epsilon)u_0}{\pi d_0\Delta x\Delta t} .
\end{equation}
The strain experienced by the drop is $\gamma=\Delta x/d_0$. Thus, the elastic modulus of the rebounding drop can be computed as
\begin{equation}
E=\frac{\sigma}{\gamma}=\frac{2m(1+\epsilon)u_0}{\pi {(\Delta x)}^2\Delta t}.
\end{equation}
using $m=2.25\times10^{-5}$ kg, we find $E=5$ MPa. A more thorough estimate using Hertz's equations\cite{landau1986theory} for two colliding elastic bodies leads to a similar estimate of $E$. Calculating the elastic modulus in this way for other impact conditions is challenging, as measuring the contact time in the absence of rebound is nontrivial. 


\section{Discussion}
In sum, our analysis presents the following picture of the drop dynamics. Upon impact, the drop experiences a large instantaneous shear at the impact point. At high enough volume fractions and impact velocities, this stress manifests itself as pockets of localised solidification embedded in the spreading liquid phase. At even higher volume fractions or shear, a larger and larger fraction of the drop solidifies after impact, but some volume at the top remains liquid.  Therefore, the shear front must be dissipating as it moves upward, and the stress falls below the critical stress for shear thickening before the entirety of the drop is solidified. Moreover, at the highest impact velocity, the  drop rebounds, and the coefficient of restitution allows us to estimate the elastic modulus of the shear jammed solid, $E=5$ MPa. Thus, our drop impact experiments provide a unique window to observe shear jamming \emph{as it occurs}, and give rise to a number of questions about the nature of both the shear jamming transition and the resulting jammed solid.

The occurrence of localised solidification coincides with the appearance of weak shear thickening in our bulk rheology data. The fact that these  solidified patches vanish over tens of milliseconds is strong evidence that they are regions of high viscosity embedded in a lower-viscosity fluid phase. Recent rheological studies using boundary stress measurements (BSM) have reported finite regions of enhanced stress in silica suspensions\cite{Rathee2017,Rathee_JRheol,rathee2022structure}. In these works, Rathee et al.\cite{Rathee2017,Rathee_JRheol,rathee2022structure} argued that the transition from shear thickening to shear jamming is governed by the growing size of such localised shear jammed regions. Our observations of transient localized solidification are thus striking visual evidence of such a mechanism. Further spatially resolved stress measurements performed on impacting drops\cite{ChengStressReview} could provide more information on the nature of localised solidification in \NewEdits{free-surface systems}.

\NewEdits{In the bulk solidification regime, the coexistence of liquid and solid regions is a result of shear traveling upward from the impact point, and simultaneously dissipating due to the high suspension viscosity. Although recent studies of low-viscosity Newtonian fluids have established the velocity and pressure fields within an impacting drop\cite{philippi_2016_newtonian,gordillo_force_2018}, they are not applicable in case of high-viscosity systems. The highly non-Newtonian nature of colloidal suspensions adds further complexity, as the fluid viscosity is a strong function of the applied shear. Numerical work investigating transient shear might be a useful next step to uncover the mechanism of dissipating shear fronts. Though challenging, measurements of the flow inside the drop via methods such as particle tracing, would provide key information about the flow field in  an impacting colloidal drop. }

The nature of the upward-travelling front raises a number of interesting questions. Before the front reaches the top, the speed of the drop apex $u^\ast$ is identical to $u_0$ [Fig. \ref{solid}b], and the curvature of the top portion is the same as it was before impact. This confirms that the information of the impact event reaches the top portion only with the front, thus establishing that it is a solidification front. Why the speed of this front is constant along the drop surface is an intriguing question. One would expect a shear front to travel through the bulk of the drop, upward from the region in contact with the substrate. Given the visual nature of our measurements on an opaque drop, we can naturally observe this front only on the surface. The most likely explanation, therefore, is that the front we measure is this bulk shear front after it interacts with the drop boundary. 

Our experiments are especially well-positioned to capture such a front due to the  free-surface conditions here that are absent in other studies of shear fronts\cite{waitukaitis2012impact,han2016high,peters2016direct,HanJaegerShearFronts,RomckePetersPRF2021}. Past work has established that shear fronts in dense suspensions are not a result of densification, and their velocity is set by the external driving speed\cite{HanJaegerShearFronts}. \NewEdits{We emphasize that the drop impact system has the unique capability to study complex fluid flow under large and instantaneous shear. As we demonstrate above, the fluid responds to this shear over a few hundred microseconds. As typical steady state rheometers apply stress to the fluid over much longer timescales, shear rates above $\dot\gamma_c$ are inaccessible. Thus, the instantaneous nature of shear in drop impact enables us to access fluid behaviour under these high shear rates.} The dependence of the front speed on $\dot\gamma_{impact}-\dot\gamma_c$ in our experiments [Fig.\ \ref{front}f] suggests that the suspension properties near the shear jamming transition are governed by the distance from the onset of shear thickening. This is consistent with measurements in static jamming, where material properties depend on the distance from the critical point\cite{LiuNagelJamming}.  The functional form of this dependence potentially contains insights into the nature of the shear jamming transition. Numerical work exploring the impact of suspension drops, although incredibly challenging due to the strong role of hydrodynamics in colloidal systems,  might provide crucial information in this respect. Unfolding the physics of these fronts will not only extend constitutive models for complex fluid rheology to much higher stress regimes, but will also help us understand more about the nature of the shear jamming transition.





Due to the transient nature of the shear jammed state, characterising the jammed solid created after impact is challenging. Using the coefficient of restitution of the rebounding drop, we were able to estimate the elastic modulus of the solid phase, $E=5$ MPa. As rebound only occurred at one impact velocity, how the elastic properties of shear jammed drops are controlled by the impact conditions remains obscure. The use of superhydrophobic substrates promotes rebound, even in Newtonian liquid drops\cite{Quere}. Further colloidal drop impact experiments on superhydrophobic surfaces could extend the parameter space where drops rebound, and thus provide the information essential to understand what controls the properties of this elastic state. Numerous other properties of the shear jammed solid are of interest: When and how would such a solid fracture? How broad is its linear elastic regime? How do these properties compare to those of static jammed solids? 
 
In conclusion, we conduct highly time-resolved drop impact experiments and systematically probe suspension flow ranging from Newtonian-like to shear jamming.  We show that the impact behaviour in the spreading regime can be quantitatively understood via an effective viscosity framework, and that the solidification behaviours at high $\phi$ and $u_0$ are direct consequences of shear jamming. The free-surface geometry in our system provides direct visual information on how the shear jamming transition occurs, both in parameter space and in time. Shear jamming occurs via a solidification front, the speed of which is set by how far into the shear thickening regime the applied shear rate is. Furthermore, we see this transition occur via a localised solidification regime that cannot be observed via bulk measurements.  We believe that drop impact is a powerful experimental tool to investigate macroscopic properties of complex fluids, and provides information that compliments the data from bulk rheometry. 







\section{Methods}
\noindent
\textbf{Colloidal sample preparation}\\
We fabricated silica spheres in our lab using the St\"ober\cite{stober1968,zhang2009} synthesis method. The particle size was determined by the number of feeds: we performed 14 feeds after the initiation of the reaction, resulting in particles with a diameter of $830 \pm 20$ $nm$. The reaction mixture was centrifuged and re-suspended in ethanol 3 times; the suspension was then gravity separated to improve monodispersity. The  particles were then imaged on the Hitachi S4800 Scanning Electron Microscope [Fig. \ref{rheology}a]. The particle size was characterized by measuring the diameter of a representative sample of 100 particles in ImageJ, and the polydispersity reported is the standard deviation in particle size. 

A concentrated stock suspension of the silica spheres was prepared in water (with no surfactant), and the weight fraction was measured by drying $100$ $\mu L$ of the stock suspension. The density of  silica ($2$ $g/cm^3$) was used 
to convert weight fractions into volume fractions. Dilutions were then performed to prepare samples of desired volume fractions. The uncertainty in volume fractions reported is $0.5\%$ $(0.005)$ or less, determined by repeated measurements. When not in use, all the sample tubes were sealed using Parafilm and stored in a refrigerator to minimize evaporation and contamination. 

\noindent
\subsection{Experimental setup}
We used Fisherbrand plain glass slides as the hydrophilic impact substrate. The slides were cleaned using a 2.5M solution of NaOH in ethanol and water to remove organic impurities. A micropipette was used to form colloidal drops. The micropipette was mounted on a vertically moving pipette holder to vary impact velocities. We used $15$ $\mu L$ of fluid to obtain drops of  $3.0 \pm 0.1$ $mm$ diameter. The setup was enclosed in a humidity chamber with the relative humidity maintained between 70--80\% using a saturated solution of NaCl in water, and the humidity was monitored in real time during experiments. Before every trial of the impact experiments, a vortex mixer was used to re-disperse the sample, ensuring that it was consistently well-mixed. 

The impacting drops were backlit using a white LED light, and filmed using two high-speed cameras. The first camera, a Phantom V2512, captured the side-view of the impacting drop at 100,000 frames per second. The second camera, a Phantom V640L, filmed at 20,000 fps. It was tilted at an angle of $15^\circ$ to gather information on how the impact affected the top surface of the drop. The experiment was repeated at least 5 times for each impact condition to ensure reproducibility. 

\noindent
\textbf{Rheological studies} \\
Stress-controlled rheological measurements were performed on the colloidal samples over $0.09 \le\phi\le 0.50$. The measurements were done on a TA Instruments Discovery HR-2 rheometer at room temperature ($\sim 21^\circ C$) using the cone-plate geometry with 40 mm diameter and a $1^\circ$ cone angle. The truncation gap was $25$ $\mu m$. We covered the edges of the samples with a microscope immersion oil to minimize evaporation. The samples were pre-sheared to remove effects of shear history.

\noindent
\textbf{Data analysis}\\
All high-speed videos were background-divided and analysed using ImageJ. The plots were made using python, and all errors reported are standard deviations over multiple trials. The maximum drop spread $d_{max}$ was determined by locating the frame in the impact timeseries where the extent of the spreading drop was the greatest. 
The height of the tallest point on the drop relative to the substrate, $h$, was measured for each frame in the image sequence. The minimum height $h_{min}$ was defined as the drop height at the crossover point between the decreasing and the plateau regimes in the $h$ vs. $t$ plot. The time of first observation of $h_{min}$, measured since the impact event, was defined as $t^\ast$. The slope of the linearly decreasing regime in the $h$ vs. $t$ plot was defined as $u^\ast$. To calculate the coefficient of restitution, the speed of the drop before impact $u_0$, and the speed after rebound, $u_f$ were computed using several frames of the image sequence. The coefficient of restitution was then computed as $\epsilon=u_f/u_0$. 

\begin{figure}
\begin{centering}
\includegraphics[width=\textwidth]{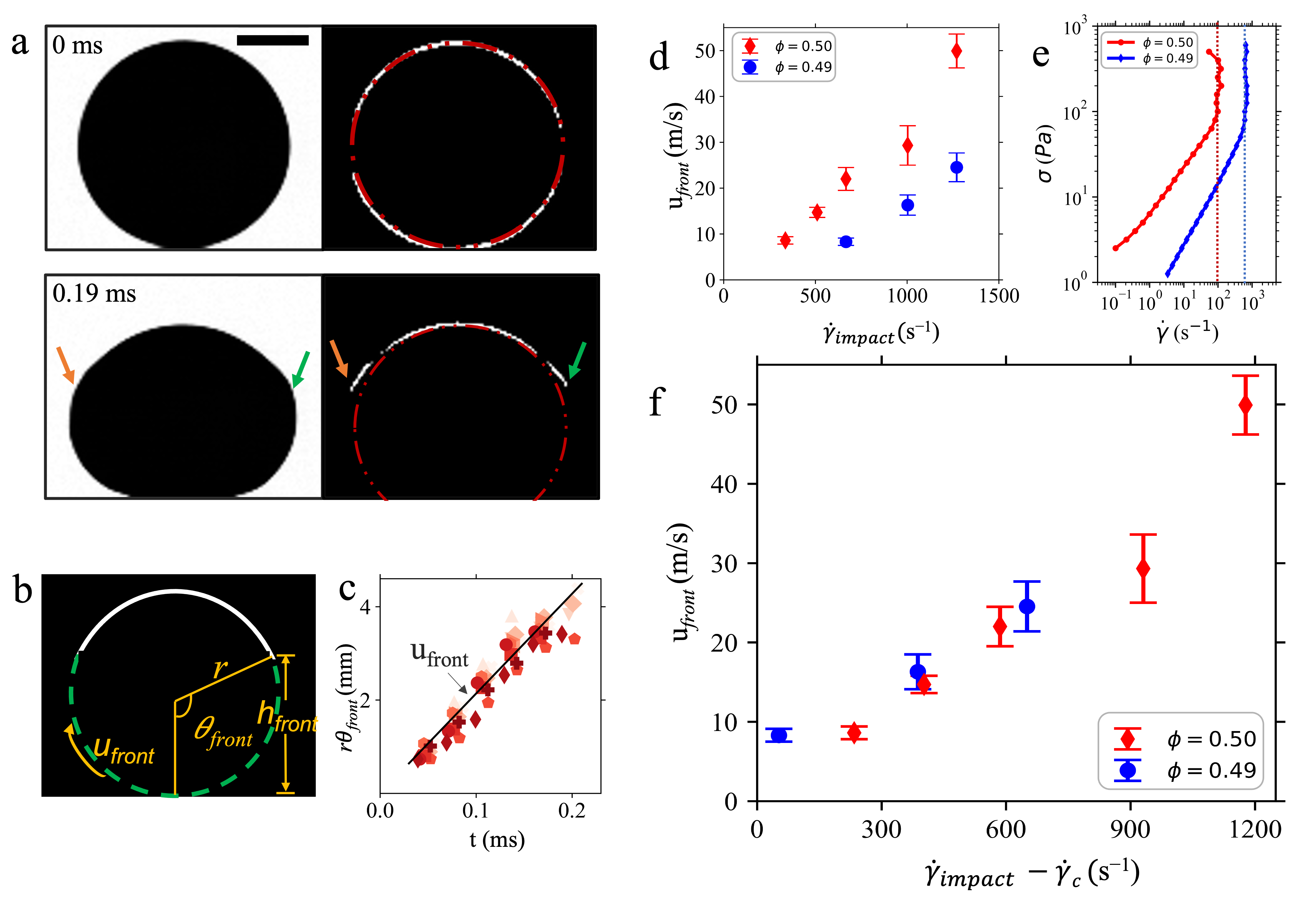} \caption{\textbf{Dynamics of the solidification front.} \textbf{a} Timeseries of a $\phi=0.50$ drop impacting at $u_0=$ 2.0 m/s [SI video 6 \cite{SI}]. Right panels are images obtained by subtracting consecutive frames, so that the edge of the solidification front is highlighted (shown by arrows).The red circle indicates the drop profile before impact. Even at 0.19 ms, the portion of the drop above the front maintains its pre-impact curvature.  Scale bar is 1 mm.~\textbf{b} Schematic of a subtracted image of the moving solidification front, outlining relevant parameters. The height, $h_{front}$, of the edge of the white outline gives the location of the front, which is then converted to $r\theta_{front}$ using the spherical geometry.~\textbf{c} Example datasets of $r\theta_{front}$ vs. $t$  for $\phi=0.49$ and $u_0=3$ $m/s$. $r\theta$ vs. $t$ is a straight line, the slope being the front speed along the surface, $u_{front}$.~\textbf{d} $u_\textit{\scriptsize{\mbox{front}}}$ plotted against $\dot\gamma_{impact}$.~\textbf{e} High- $\phi$ bulk rheological data from Fig.\ref{rheology}b re-plotted as shear stress vs. shear rate. Dotted lines indicate the onset shear rates $\dot\gamma_c$ for shear thickening. ~\textbf{f} The $u_\textit{\scriptsize{\mbox{front}}}$ data for $\phi=0.49$ and $\phi=0.50$, when plotted against $\dot\gamma_{impact}-\dot\gamma_c$, collapses on a single curve. }
\label{front}
\end{centering}
\end{figure}

To calculate the speed of the upward-moving front, the side-view impact videos recorded at 100,000 fps were used. For every frame of the image sequence, the pixel-wise difference between consecutive frames was taken in ImageJ, so that only the elements that changed between consecutive frames (corresponding to the location of the moving front) were highlighted [SI video 7 \cite{SI}]. This enabled us to locate the jamming front with a time uncertainty of 10 $\mu$s. The images were then adjusted for brightness and contrast to enhance the moving front. The vertical height $h_\textit{\scriptsize{\mbox{front}}}$ of the disturbance from the impact substrate was measured for each frame of the image sequence, until the front was no longer visible. For every high-speed video, the left and right half of the drop were separately analyzed to obtain two datasets for $h_\textit{\scriptsize{\mbox{front}}}(t)$. In order to convert $h_\textit{\scriptsize{\mbox{front}}}$ to the position along the drop surface, $r\theta_\textit{\scriptsize{\mbox{front}}}(t)$, we approximated the drop profile as a circle of radius $r= 1.5$ $mm$ (disregarding the slight deviation from spherical shape during front propagation), and used the relation $h_\textit{\scriptsize{\mbox{front}}}(t)=r(1-\cos\theta_\textit{\scriptsize{\mbox{front}}}(t))$, such that $\theta\textit{\scriptsize{\mbox{front}}}(0) = 0$ at the impact point, to obtain the angle $\theta_\textit{\scriptsize{\mbox{front}}}(t)$. A line was then fit to the $r\theta_\textit{\scriptsize{\mbox{front}}}$ vs.\ time plots, and the slope, averaged over  the two halves of the drop and several movies for each impact condition [Fig.\ \ref{front}c],  was reported as $ u_\textit{\scriptsize{\mbox{front}}}$ with error bars indicating the standard deviation.




\begin{addendum}
 \item [Acknowledgements]  We thank Jeff Richards, Sid Nagel, and Xiang Cheng for useful discussions. This work was supported by the National Science Foundation under award number DMR-2004176.  This work made use of the EPIC facility of Northwestern University's NUANCE Center, which has received support from the SHyNE Resource (NSF ECCS-2025633), the IIN, and Northwestern's MRSEC program (NSF DMR-1720139). We thank the Richards Lab at Northwestern University for the use of their rheometry facilities.
 \item[Competing Interests] The authors declare that they have no
competing financial interests.
 \item[Correspondence] Correspondence and requests for materials
should be addressed to MMD.

(michelle.driscoll@northwestern.edu).
\item[Author contributions] SA contributed to the conception of the work, experimental design, and data acquisition and analysis.  PS contributed to data interpretation and analysis and drafted the manuscript. MMD contributed to the conception of the work, data analysis and interpretation, and drafted the manuscript.

\end{addendum}

\newpage

\bibliography{Bib_nature}
\bibliographystyle{naturemag}

\newpage

\end{document}